\documentclass[structabstract]{aa}  
\usepackage{graphicx}
\usepackage{txfonts}
\usepackage[colorlinks=true,linkcolor=blue,citecolor=blue,urlcolor=black]{hyperref}
\usepackage{natbib}
\bibpunct{(}{)}{;}{a}{}{,}

\begin{document}
   \title{The inner circumstellar disk of the UX~Ori star V1026~Sco
     \thanks{Based on observations made with ESO telescopes at the La
       Silla Paranal Observatory under programme IDs 083.D-0224(C),
       083.C-0236(A), 087.C-0013(A) and 073.A-9014(A)} }

   \titlerunning{The circumstellar disk of the UX~Ori star V1026~Sco}

   \author{
     J. Vural \inst{1} \thanks{Member of the International Max
       Planck Research School (IMPRS) for Astronomy and Astrophysics
       at the Universities of Bonn and Cologne}, 
     A. Kreplin \inst{1},
     M. Kishimoto \inst{1}, 
     G. Weigelt \inst{1}, 
     K.-H. Hofmann \inst{1},
     S. Kraus \inst{2},
     D. Schertl \inst{1},
     M. Dugu\'e \inst{3},
     G. Duvert \inst{4},
     S. Lagarde \inst{3},
     F. Massi \inst{5} }

   \authorrunning{J. Vural et al.}

   \institute{ 
     Max-Planck-Institut f\"ur Radioastronomie, Auf dem H\"ugel 69,
     53121 Bonn, Germany
     \and University of Exeter, Astrophysics group, Physics Building,
     Stocker Road Exeter, EX4 4QL, UK
     \and Laboratoire Lagrange, UMR7293, Universit\'e de Nice
     Sophia-Antipolis, CNRS, Observatoire de la C\^ote d'Azur, 06300
     Nice, France 
     \and UJF-Grenoble 1 / CNRS-INSU, Institut de Plan{\'e}tologie et
     d'Astrophysique de Grenoble (IPAG) UMR 5274, Grenoble, F-38041,
     France
     \and INAF - Osservatorio Astrofisico di Arcetri, Largo E. Fermi,
     5, 50125 Firenze, Italy 
     }

   \date{Received ...; accepted ...}

 
  \abstract
  {The UX Ori type variables (named after the prototype of their
    class) are intermediate-mass pre-main sequence objects. One of the
    most likely causes of their variability is the obscuration of the
    central star by orbiting dust clouds.}
 {We investigate the structure of the circumstellar environment of the
   UX~Ori star V1026~Sco (HD~142666) and test whether the disk
   inclination is large enough to explain the UX~Ori variability. }
 {We observed the object in the low-resolution mode of the
   near-infrared interferometric VLTI/AMBER instrument and derived
   {\textit H}- and {\textit K}-band visibilities and closure
   phases. We modeled our AMBER observations, published Keck
   Interferometer observations, archival MIDI/VLTI visibilities, and
   the spectral energy distribution using geometric and
   temperature-gradient models. }
 {Employing a geometric inclined-ring disk model, we find a ring
   radius of $0.15 \pm 0.06$~AU in the {\textit H} band and $0.18 \pm
   0.06$~AU in the {\textit K} band.  The best-fit
   temperature-gradient model consists of a star and two concentric,
   ring-shaped disks. The inner disk has a temperature of
   $1257^{+133}_{-53}$~K at the inner rim and extends from $0.19 \pm
   0.01$~AU to $0.23 \pm 0.02$~AU. The outer disk begins at
   $1.35^{+0.19}_{-0.20}$~AU and has an inner temperature of
   $334^{+35}_{-17}$~K. The derived inclination of
   $48.6^{+2.9}_{-3.6}${\degr} approximately agrees with the
   inclination derived with the geometric model ($49 \pm 5\degr$ in
   the {\textit K} band and $50 \pm 11\degr$ in the {\textit H}
   band). The position angle of the fitted geometric and
   temperature-gradient models are $163 \pm 9\degr$ ({\textit K}~band;
   $179 \pm 17\degr$ in the {\textit H}~band) and
   $169.3^{+4.2}_{-6.7}$\degr, respectively.}
  {The narrow width of the inner ring-shaped model disk and the disk
   gap might be an indication for a puffed-up inner rim shadowing
   outer parts of the disk. The intermediate inclination of
   $\sim$50{\degr} is consistent with models of UX~Ori objects where
   dust clouds in the inclined disk obscure the central star.}

   \keywords{Stars: individual: V1026~Sco, Stars: pre-main sequence,
     formation, circumstellar matter, Techniques: interferometric }

   \maketitle
%

\section{Introduction}

The UX Ori (UXOr) phenomenon of Herbig Ae/Be stars (HAeBes) is
attributed to obscuration by circumstellar dust in an inclined disk
\citep{1994grithe,1997natgri,2001grikoz,2003dulvan} or unsteady
accretion \citep{1999hershe}. The Herbig Ae star \object{V1026~Sco}
(HD~142666) has a spectral type of A8Ve \citep{2003domdul} and is
classified as a UX~Ori object \citep{1998meewae}. The UX~Ori
variability has been confirmed by
\citet{2009zwikal}. \citet{2003domdul} and \citet{2005vanmin} report
distances of 116~pc and $145 \pm 43$~pc, respectively. We adopt the
Hipparcos-based measurement of 116~pc and the associated parameters
for our work. The object V1026~Sco shows large, non-periodic
\citep{2005lecnit} brightness variations ($>$1.2~mag) and a
pulsational variability on the milli-magnitude level
\citep{2009zwikal}. It reddens with decreasing apparent magnitude
\citep{1998meewae}. These authors suggest that dense dust clouds in an
inclined disk cause the stellar reddening. \citet{2013alewad} report
on the magnetic properties of V1026~Sco (and several other Herbig
Ae/Be stars). The object V1026~Sco belongs to the Meeus group IIa
\citep{2010juhbou} and might, therefore, have a self-shadowed
disk. The stellar parameters \citep{2003domdul} of V1026~Sco are
listed in Table~\ref{tabpro}.
By modeling the spectral energy distribution (SED), \citet{2003domdul}
found that the circumstellar disk of V1026~Sco has an inclination of
approximately 55\degr. \citet{2005monmil} have performed Keck
Interferometer (KI) measurements of V1026~Sco and found an inner disk
diameter of 2.52~mas (0.29~AU at 116~pc).
In a recent publication, \citet{2013schrat} have reported mid- and
near-infrared interferometric observations (archival MIDI/VLTI \& IOTA
data), and performed radiative transfer modeling of V1026~Sco, and
derived a disk structure with a gap from 0.35~AU to 0.80~AU.

In this paper, we analyze the circumstellar environment around
V1026~Sco by taking new interferometric near-infrared (NIR) VLTI/AMBER
and archival mid-infrared (MIR) VLTI/MIDI measurements into
account. We describe our observations and the data reduction in
Sect.~\ref{kapobs}. The modeling is presented in Sect.~\ref{kapmod},
and our results are discussed in Sect.~\ref{kapdis}.

\begin{table}[t]
\caption{The adopted stellar parameters of HD142666.}     
\label{tabpro}   
\centering                       
\begin{tabular}{rcl}  
\hline\hline
Parameter        & Value                         \\ \hline
spectral type    & A8Ve \tablefootmark{a}        \\
age [Myr]        & $6.0 \pm 1.5$ \tablefootmark{b} \\
distance [pc]    & 116                           \\
$M_* [M_\odot]$   & 1.8                           \\
$L_* [L_\odot]$   & 11                            \\
$T_* [\mathrm K]$& 8500                          \\
$log(\dot M [M_\odot \mathrm{yr}^{-1}])$ 
                 & $-6.73 \pm 0.26$ \tablefootmark{c} \\
\hline
\end{tabular}
\tablefoot{The values are taken from \citet{2003domdul} unless
  otherwise noted. The error bars are shown where available.
  \citet{2003domdul} estimate the uncertainty of the luminosity to be
  $\pm50\%$ (due to the Hipparcos distance error) and the mass
  uncertainty to be 5 -- 10\%. Other authors find slightly different
  parameters for the alternative distance of $145 \pm 20$~pc. For
  example \citet{2013alewad} derived $5.0^{+1.6}_{-1.1}$~Myr,
  $2.15^{+0.20}_{-0.19} M_\odot$, $27.5^{+7.9}_{-7.1} L_\odot$, and
  $7900 \pm 200$~K. Other references:
  \tablefoottext{a}{\citet{1998meewae}},
  \tablefoottext{b}{\citet{2012folbag}},
  \tablefoottext{c}{\citet{2011mencal}}. }
\end{table}

\section{Observation and data reduction} \label{kapobs}
\begin{figure}[t]
  \includegraphics[angle=-90,width=0.4\textwidth]{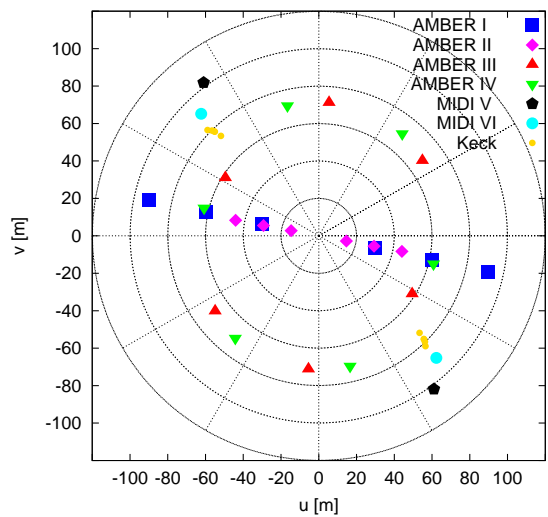}
  \caption{The uv coverage of all interferometric measurements used
    (AMBER, MIDI, Keck; see Table~\ref{tabobs1}). }
  \label{figuv}
\end{figure}

We observed V1026~Sco on three different nights with the NIR
three-beam VLTI/AMBER instrument \citep{2007petmal}. The observations
were performed in low-resolution mode (spectral resolution
$R=30$). Table~\ref{tabobs1} lists the observational parameters. The
uv coverage of our 2009 and 2011 AMBER observations with the published
KI and archival MIDI observations of V1026~Sco used in this study are
shown in Figure~\ref{figuv}.

We reduced the AMBER data with {\it
  amdlib-3.0.2}\footnote{\url{http://www.jmmc.fr/data_processing_amber.htm}}
\citep{2007tatmil,2009cheutr}. To improve the calibrated visibility,
we processed only 20\% of the frames (object and calibrator) with the
best fringe signal-to-noise ratio \citep{2007tatmil}. In addition, we
equalized the histograms of the optical path differences (OPD) of the
calibrator and the object data, because different histograms (due to
OPD drifts caused, for example, by errors of the OPD model) can lead
to visibility errors. Histogram equalization can reduce these
visibility errors. This method is described in detail in
\citet{2012krekra}.

\begin{figure*}
  \centering
  \includegraphics[width=\textwidth]{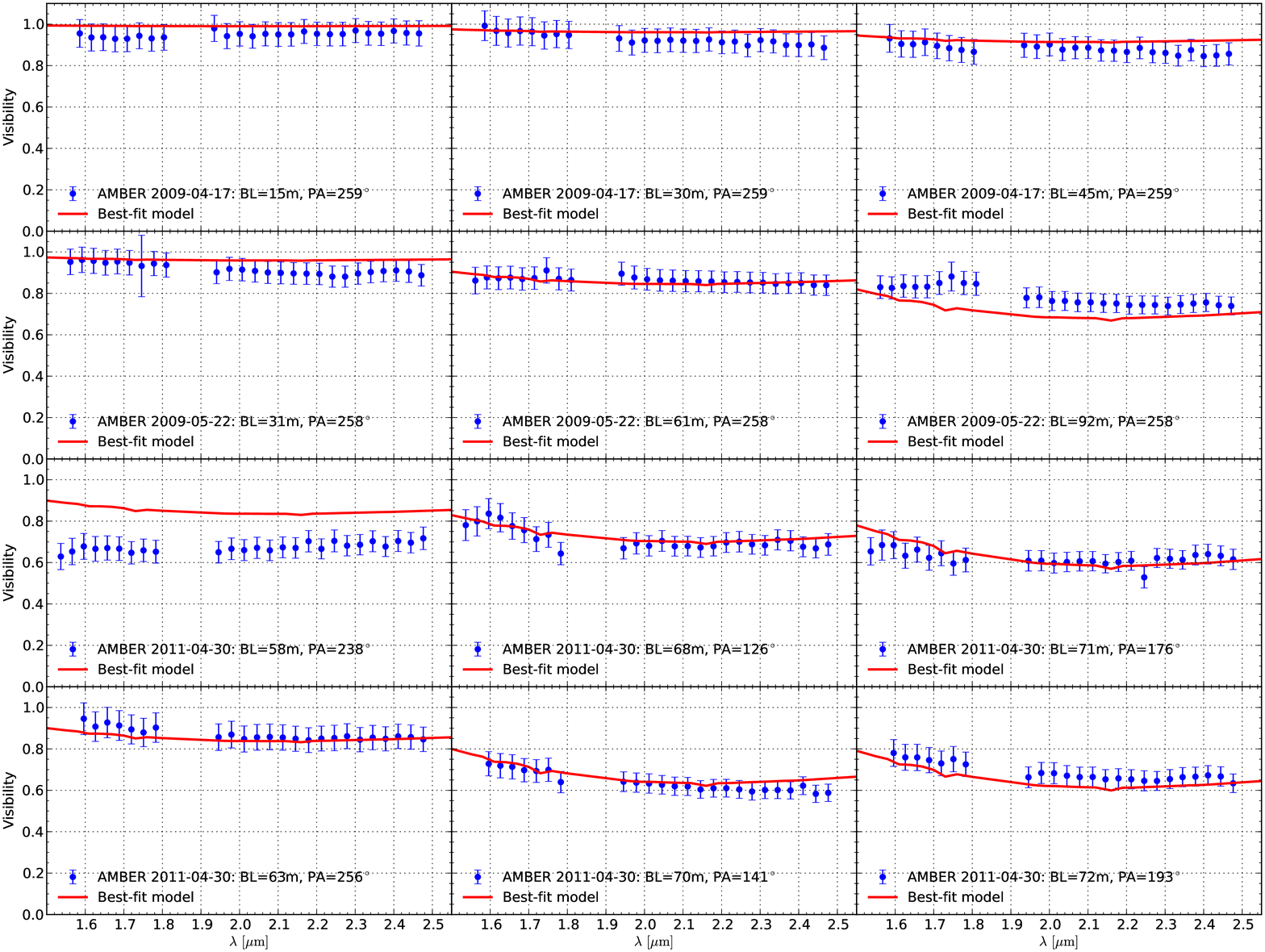}
  \caption{AMBER visibilities (see Table~\ref{tabobs1}) and
    temperature-gradient model B$_3$: The panels show the
    wavelength-dependent {\textit H} and {\textit K} band visibilities
    of our AMBER observations. Each panel displays one of the three
    baselines of each measurement (nights I-IV,
    cf. Tab.~\ref{tabobs1}). The red line indicates the corresponding
    best-fit temperature-gradient model curves (model~B$_3$ in
    Table~\ref{tabres}) in all plots.}
  \label{figtgma} 
\end{figure*}

\begin{figure}
  \centering
  \includegraphics[width=0.5\textwidth]{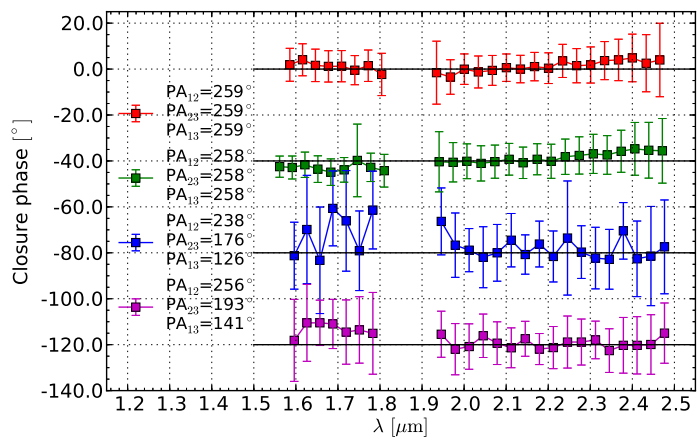}
  \caption{Wavelength dependence of the closure phases of all AMBER
    measurements from Table~\ref{tabobs1}. The curves are offset from
    each other by 40\degr. The respective zero line is indicated as a
    solid horizontal line.}
  \label{figcp}
\end{figure}

We were able to extract {\textit H} and {\textit K} band visibilities
(cf. Fig.~\ref{figtgma}). Within the error bars, the obtained closure
phase is zero for all nights (Fig.~\ref{figcp}), which is consistent
with a centro-symmetric brightness distribution.

\begin{figure*}
  \centering
  \begin{minipage}{0.33\textwidth}
    \includegraphics[width=\textwidth]{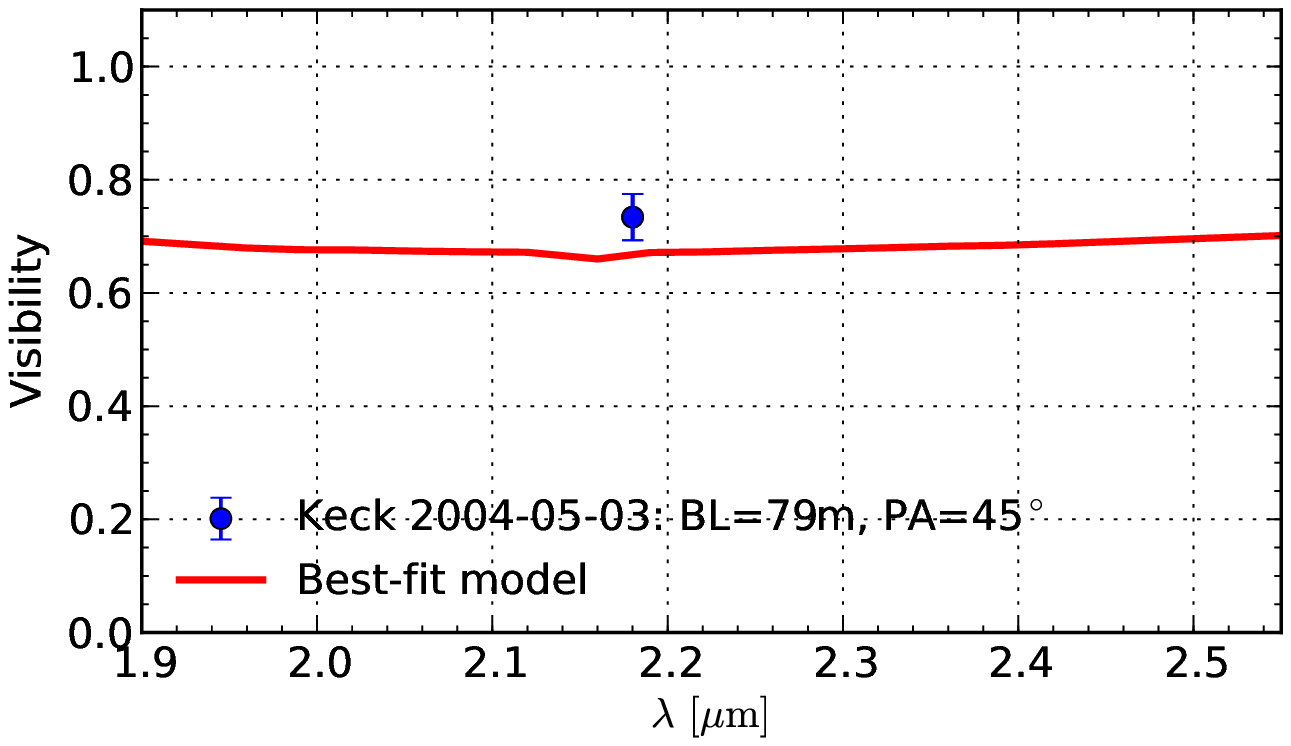}
  \end{minipage}
  \begin{minipage}{0.66\textwidth}
    \includegraphics[width=\textwidth]{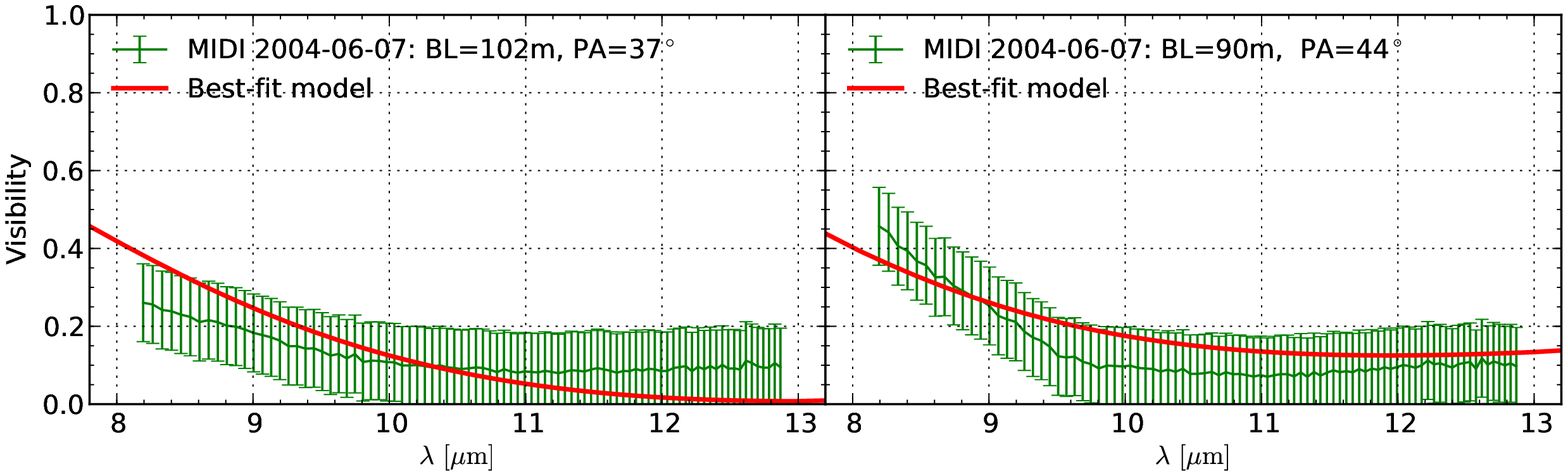}
  \end{minipage}
  \caption{Keck and MIDI observations and temperature-gradient model
    B$_3$. {\it Left:} Keck visibility. {\it Right:} MIDI
    visibilities.  }
  \label{figtgmm} 
\end{figure*}

We also used archival low-resolution data obtained with the MIR
two-beam combiner MIDI \citep{2003leigra,2013schrat}. The data (see
Table~\ref{tabobs1}) were reduced using our own IDL codes (see
Appendix A of \citealt{2011kishoe} for details), which utilize a part
of the standard software
EWS\footnote{\url{http://www.strw.leidenuniv.nl/~nevec/MIDI/index.html}}
and also implement an average over a relatively large number of frames
to determine the group-delay and phase-offset tracks with a good
signal-to-noise ratio. This is important when dealing with sub-Jy
sources, such as our target here. The MIDI visibilities are shown in
Fig.~\ref{figtgmm} (right).

\begin{table*}[htb]
  \caption{Observation log.}     
  \label{tabobs1}   
  \centering                       
  \begin{tabular}{cccccccccc}  
    \hline\hline
    \#  & Night      & Instrument & $B_\mathrm{proj}$ & PA          & Seeing    & DIT  & Calibrator & Calibrator diameter \\
        &            &            &  [m]             & [\degr]     & [\arcsec] & [ms] &            & [mas]    \\ \hline
    I   & 2009-04-17 & AMBER      & 15/30/45         & 259/259/259 & 0.55      & 200  & HD142669   & $0.256 \pm 0.018$ \tablefootmark{a} \\
    II  & 2009-05-22 & AMBER      & 31/61/91         & 258/258/258 & 0.8       & 150  & HD142669   & $0.256 \pm 0.018$ \tablefootmark{a} \\
    III & 2011-04-30 & AMBER      & 59/68/71         & 238/126/176 & 1.0       & 300  & HD138472   & $1.07 \pm 0.02$ \tablefootmark{b}   \\
    IV  & 2011-04-30 & AMBER      & 63/70/72         & 256/141/193 & 0.8       & 300  & HD138472   & $1.07 \pm 0.02$ \tablefootmark{b}   \\
    V   & 2004-06-07 & MIDI       & 102              &         37  & 0.9       & 14   & HD146791   & $3.00 \pm 0.033$ \tablefootmark{c} \\
    VI  & 2004-06-07 & MIDI       & 90               &         44  & 1.1       & 18   & HD169916   & $4.24 \pm 0.047$ \tablefootmark{c} \\
    \hline
  \end{tabular}
  \tablefoot{$B_\mathrm{proj}$ denotes the lengths of the projected
    interferometer baselines during the observations, PA the baseline
    position angles, and DIT the detector integration time for
    recording individual interferograms. The resulting uv coverage is
    shown in Fig.~\ref{figuv}. References:
    \tablefoottext{a}{\citet{2010lafmel}},
    \tablefoottext{b}{\citet{2005ricper}},
    \tablefoottext{c}{\citet{2002borcou}}.}
\end{table*}

\section{Analysis} \label{kapmod}

\subsection{Geometric modeling} \label{kapgeo}
\begin{figure*} 
  \centering
  \begin{minipage}[b]{0.49\textwidth} 
    \includegraphics[angle=-90,width=\textwidth]{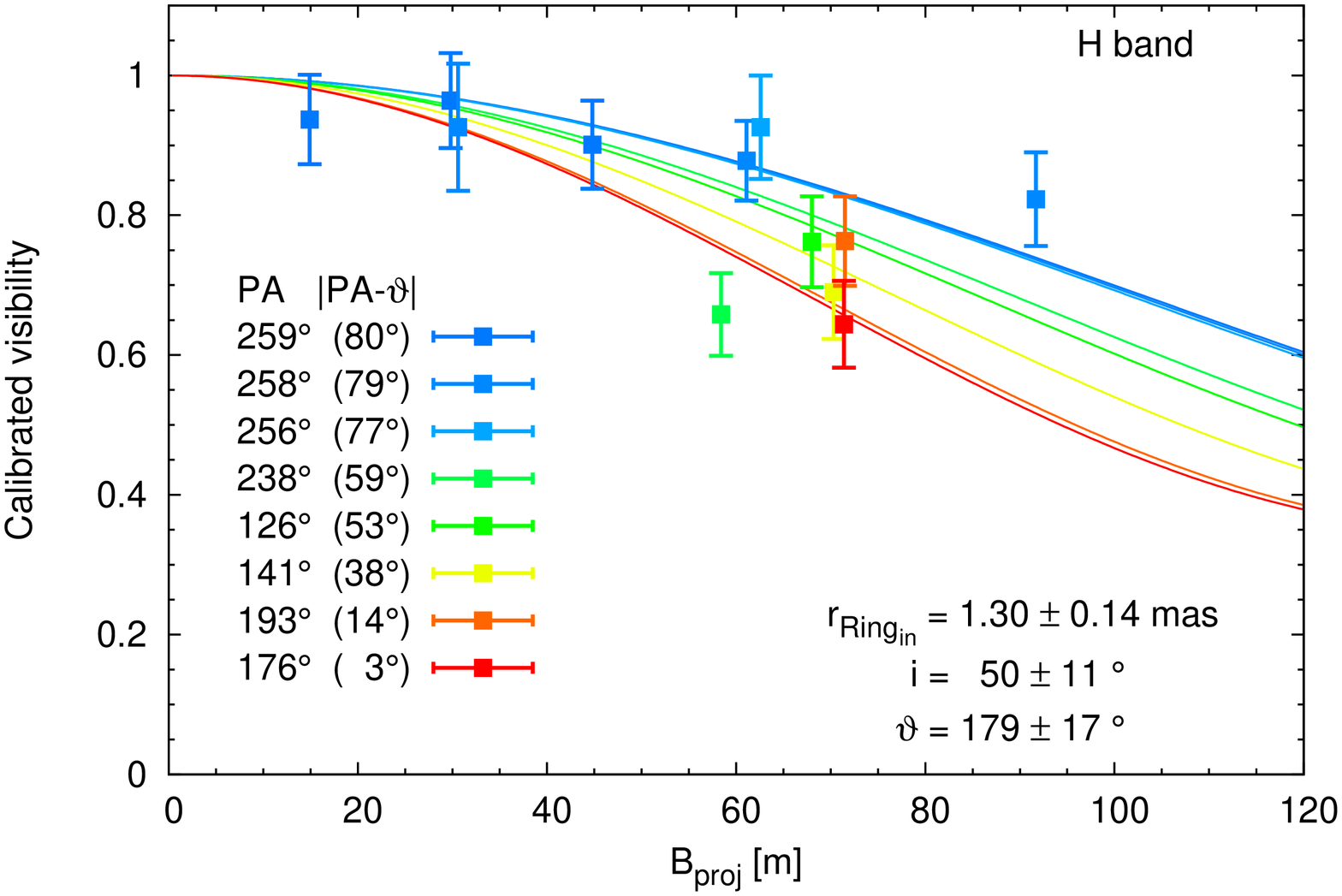}
  \end{minipage}
  \begin{minipage}[b]{0.49\textwidth} 
    \includegraphics[angle=-90,width=\textwidth]{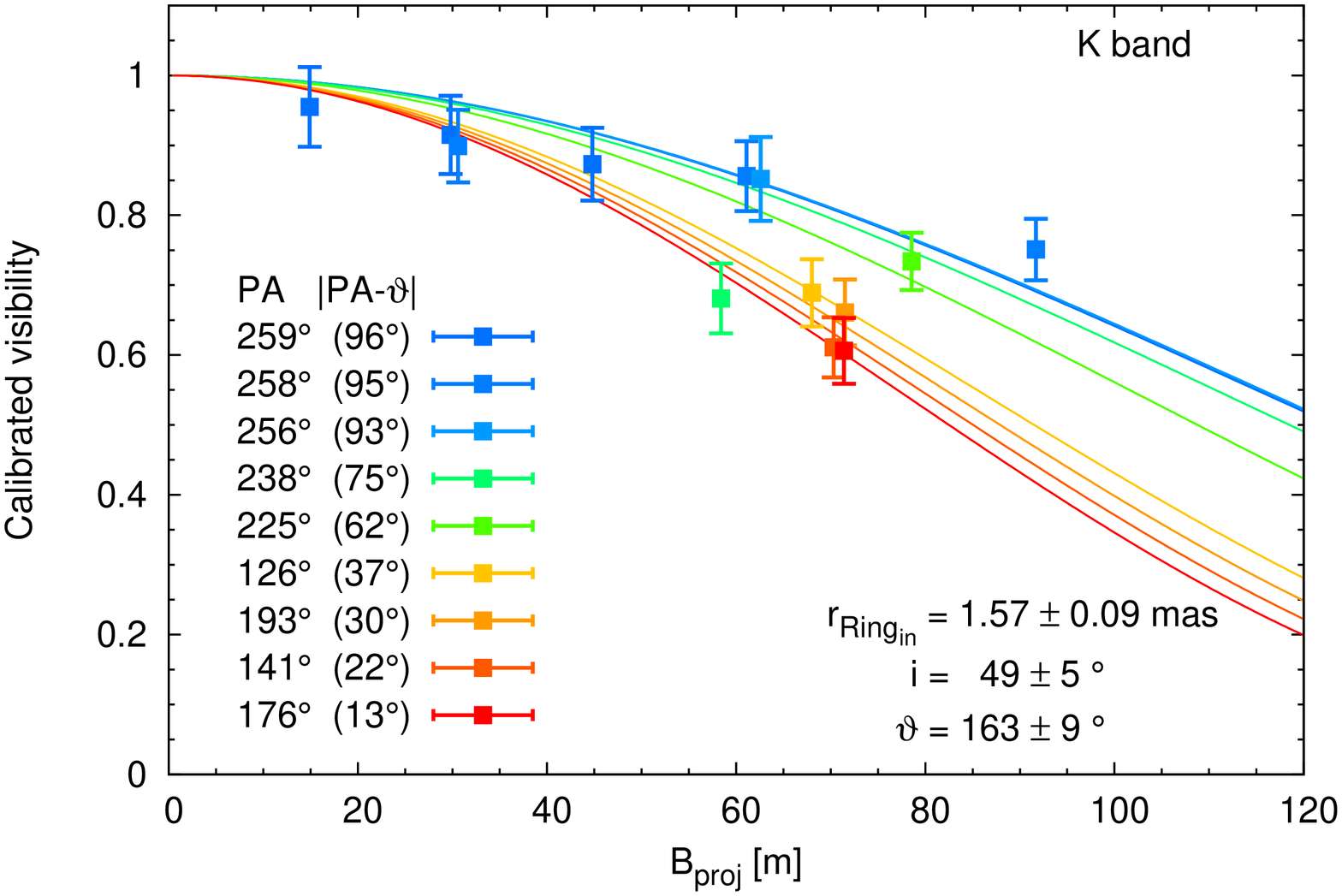}
  \end{minipage}
  \caption{Geometric inclined ring-fit models (ring width $=20\%$ of
    $r_\mathrm{ring,in}$) of the {\textit H}- (left) and {\textit
      K}-band (right) visibilities. The right plot contains our AMBER
    data and the KI measurements. The wavelengths are averaged over
    the whole respective spectral band. The model consists of the
    unresolved stellar contribution and an inclined ring (see
    Sect.~\ref{kapgeo}). We simultaneously fit all visibilities with a
    two-dimensional visibility model. The model curves are plotted for
    all position angles for which visibilities were measured. The
    color sequence (blue to red) describes the decreasing difference
    between the PA of the measurement and the disk's fitted semi-major
    axis $\vartheta$. }
 \label{figdia}
\end{figure*}

To estimate the characteristic size of the NIR emission region, we fit
geometric models to the visibilities. The model for the NIR data
consists of an unresolved stellar contribution and an inclined ring
with a width of 20\% of its inner radius $r_\mathrm{ring,in}$ (,which
is equal to the semi-major axis in the model). We averaged the
different visibility measurements (shown in Fig.~\ref{figtgma}) of the
spectral channels within the {\textit H} and {\textit K} bands to
obtain wavelength-averaged {\textit H}- and {\textit K}-band
visibilities (Fig.~\ref{figdia}). The resulting visibilities within
each band were fit with the model visibilities of the two-dimensional,
inclined star-ring model.

In the {\it K}~band, we included literature data from the Keck
interferometer (\citealt{2005monmil}, see Fig.~\ref{figuv}) in the
fit. Because of the almost constant projected baseline length and
position angle of the five KI measurements, the data points were
averaged.

\begin{figure*}[htb]
  \centering
  \begin{minipage}{0.49\textwidth}
    \includegraphics[width=\textwidth]{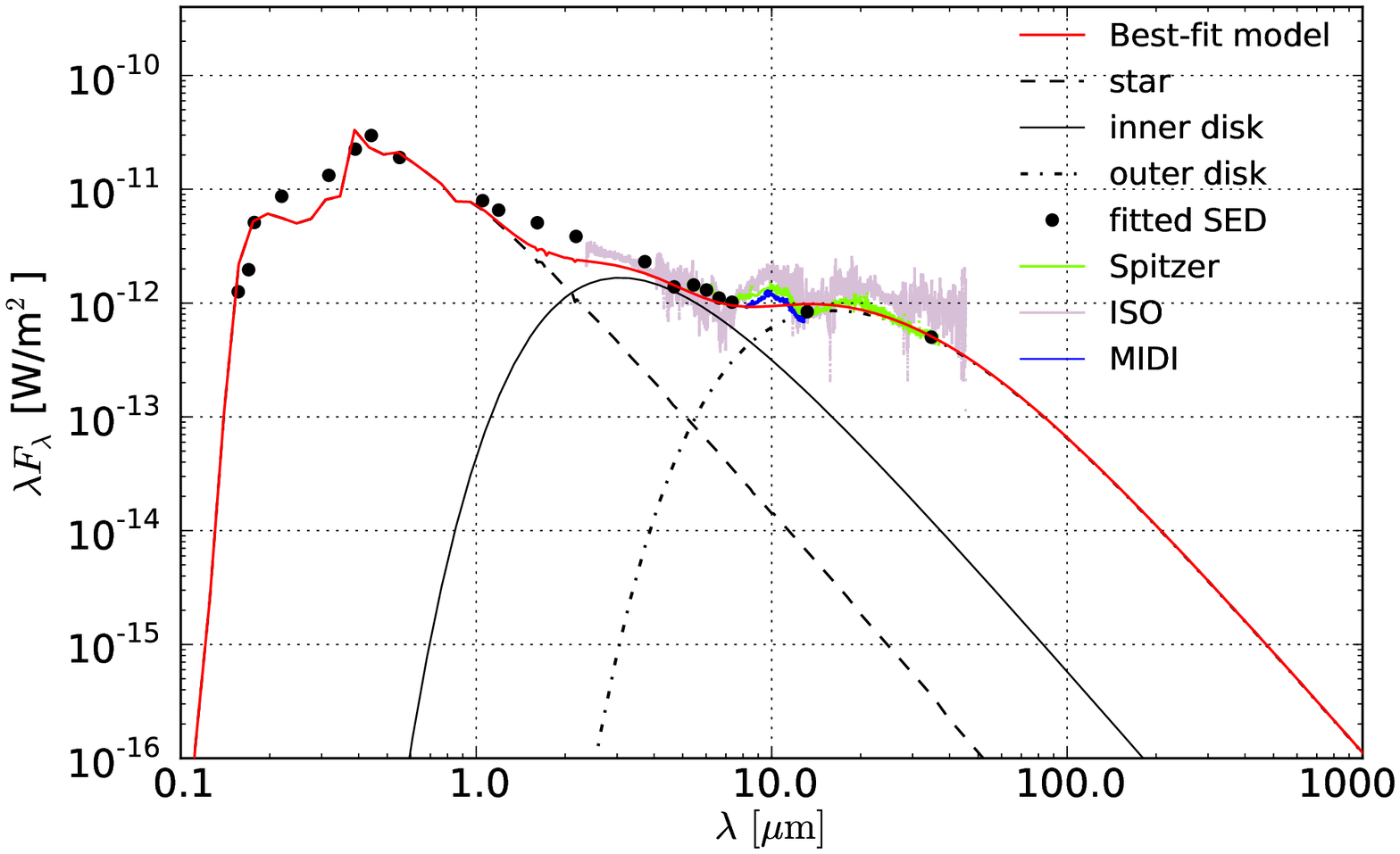}
  \end{minipage}
  \begin{minipage}{0.36\textwidth}
    \includegraphics[angle=0,width=\textwidth]{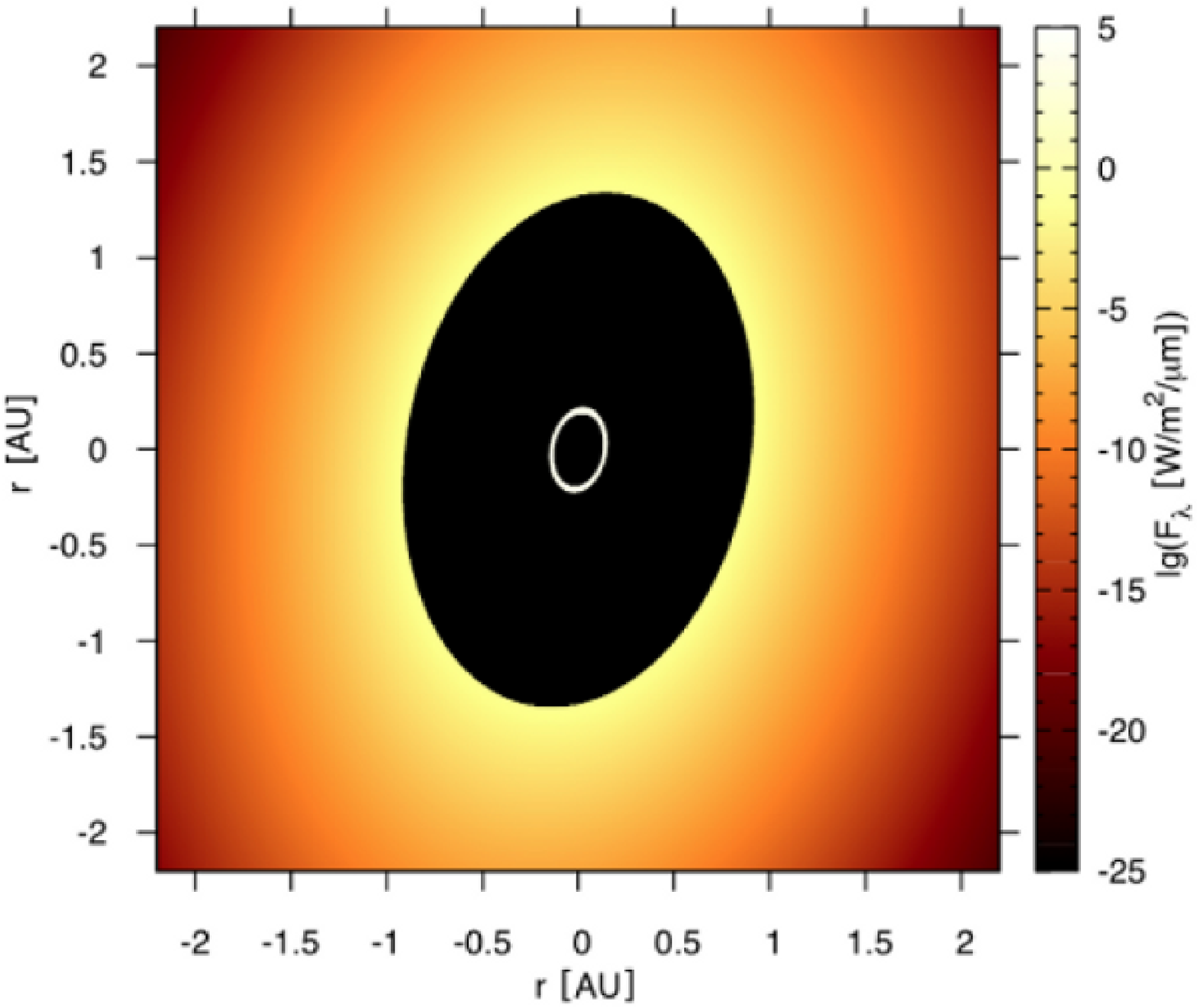}
  \end{minipage}
  \caption{{\it Left:} SED and temperature-gradient model B$_3$: The
    SED of V1026~Sco was constructed using published SED observations
    (black dots). Some of the points represent binned observations;
    the size of the error bars is on the order of the size of the
    dots. The resulting fit curve (red) consists of the stellar
    contribution (Kurucz model, black dashed line), the inner ring
    (black solid line), and the outer ring model SED (black
    dash-dotted line). {\it Right:} Two-dimensional intensity
    distribution of our best-fit model (B$_3$) at $2~\mu$m. The narrow
    bright ring is the inner ring-shaped disk in the model. The
    central star is not shown here. Please note that the intensity
    scale is logarithmic; the outer disk contributes only
    insignificantly to the NIR flux. The actual geometry of the outer
    disk remains poorly constrained because only two MIDI uv points
    exist.}
  \label{figtgms} 
\end{figure*}

To fit the visibilities, we derived the NIR flux contribution of the
star ($f_\mathrm{star}$) from our SED fit (Fig.~\ref{figtgms} left)
and obtained approximately 0.33 (\citealt{2005monmil} report 0.39) in
the {\textit K} band and 0.53 in the {\textit H} band. For the total
visibility, we obtain
\begin{equation}
  |V| = |(1-f_\mathrm{star})V_\mathrm{disk} + f_\mathrm{star}V_\mathrm{star}| \;,
\end{equation}
where $f_\mathrm{star}+f_\mathrm{disk}=1$ and the unresolved star has
$V_\mathrm{star}=1$.

We found a semi-major axis ($r_\mathrm{ring,in}$) of $1.30 \pm
0.14$~mas (or $0.15 \pm 0.06$~AU for a distance of 116~pc) in the
{\textit H} band and $1.57 \pm 0.09$~mas (or $0.18 \pm 0.06$~AU) in
the {\textit K} band (Fig.~\ref{figdia}). All fitted parameters are
listed in Table~\ref{tabrad}. In the {\textit H} band, the inclination
angle $i$ (angle between the system axis and the viewing direction) is
$50 \pm 11\degr$, and the position angle $\vartheta$ of the semi-major
axis of the disk is $179 \pm 17\degr$. In the {\textit K} band, we
derived $i=49 \pm 5\degr$ and $\vartheta=163 \pm 9\degr$,
respectively.

\begin{table*}[thb]
  \centering  
  \caption{Parameters of the best-fit geometric model.}
  \label{tabrad}                     
  \begin{tabular}{ccccccc}  
    \hline \hline
    Band        & $r_\mathrm{ring,in}$ & $i$        & $\vartheta$  & $f_\mathrm{star}$           & $f_\mathrm{disk}$ \\ 
                & [AU]                & \degr      & \degr        &                 &                  \\ \hline
    {\textit H} & $0.15 \pm 0.06$     & $50 \pm 11$& $179 \pm 17$ & $0.53 \pm 0.10$ & $0.47 \pm 0.10$  \\
    {\textit K} & $0.18 \pm 0.06$     & $49 \pm 5$ & $163 \pm 9$  & $0.33 \pm 0.27$ & $0.67 \pm 0.27$  \\
    \hline
  \end{tabular}
  \tablefoot{The radius $r_\mathrm{ring,in}$ represents the inner
    semi-major axis of the inclined model ring; the position angle
    $\vartheta$ denotes the position angle of the semi-major axis; $i$
    is the inclination of the system axis to the line of sight,
    $f_\mathrm{star}$ the flux contribution from the star, and
    $f_\mathrm{disk}$ the flux contribution from the disk to the total
    flux (see text in Sect.~\ref{kapgeo}). The errors of the radii
    include the distance error ($116 \pm 30$~pc).}
\end{table*}

\subsection{Temperature-gradient model} \label{kaptgm}
To fit all available wavelength-dependent visibilities (AMBER, MIDI,
and KI, see Figs. \ref{figtgma}, \ref{figtgmm}) and the SED
(Fig.~\ref{figtgms}) simultaneously, we used a temperature-gradient
model. This model consists of many thin rings emitting blackbody
radiation at a local temperature $T(r)=T_0 \cdot
\left(r/r_0\right)^{-q}$, where $r_0$ denotes the inner disk radius,
$T_0$ the effective temperature at $r_0$, and $q$ the power-law
index. Other free parameters are the inclination $i$ and the position
angle $\vartheta$ of the semi-major axis of the disk. A more detailed
description of our modeling of temperature-gradient disks can be
found, for example, in \citet{2012vurkre} or \citet{2012krekra}.

We computed models for all mathematical combinations of the parameter
values listed in Table~\ref{tabpar}. We chose the model with the
lowest $\chi^2_\mathrm{red}$ value as the best-fit model. The given
error bars are $3\sigma$ errors.

We first attempted to fit the data with a model consisting of a
stellar point source (distance, $L_*$, and $T_*$ from
Table~\ref{tabpro}) and a temperature-gradient disk (model~A in
Table~\ref{tabpar}). The model has six free parameters: the inner ring
radius $r_\mathrm{in,1}$, the width of the ring $\Delta
r_\mathrm{in,1}=r_\mathrm{out,1}-r_\mathrm{in,1}$, the temperature at
the inner radius $T_\mathrm{in,1}$, the power-law index $q_1$, the
inclination $i$, and the position angle of the semi-major axis of the
disk-like object $\vartheta$. The best-fit parameters are listed in
Table~\ref{tabres}. However, no successful fit could be found that is
able to reproduce all observations simultaneously
($\chi^2_\mathrm{red}=6.2$).

Therefore, we adopted a two-component model consisting of the star
(same parameters as above) and two inclined concentric ring-shaped
disks (model~B, see Table~\ref{tabpar} and \ref{tabres}). There are
ten free model parameters: four for the inner disk ($r_\mathrm{in,1}$,
$\Delta r_\mathrm{in,1}$, $T_\mathrm{in,1}$, $q_1$), four for the
outer disk ($r_\mathrm{in,2}$, $\Delta r_\mathrm{in,2}$,
$T_\mathrm{in,2}$, $q_2$), and two for the whole disk system ($i$,
$\vartheta$). We computed all combinations of these parameters within
the parameter ranges (and for the described $N$ step values) defined
in Table~\ref{tabpar}. We first calculated the models with a rough
grid (model B$_1$) and then with finer grids (model B$_2$ and B$_3$)
around the $\chi^2_\mathrm{red}$-minimum of the previous run. In total
(for all models described in Table~\ref{tabpar}), we computed
$\sim$700~million models.

In our best-fitting model B$_3$ ($\chi^2_\mathrm{red}=2.7$, see
Figs.~\ref{figtgma},~\ref{figtgmm}, and ~\ref{figtgms}), the inner
disk spans from $0.19 \pm 0.01$~AU to $0.23 \pm 0.02$~AU (with a
temperature of $1257^{+133}_{-53}$~K at the inner radius
$r_\mathrm{in,1}$) and the outer disk between
$1.35^{+0.19}_{-0.20}$~AU and $>4.3$~AU ($334^{+35}_{-17}$~K at
$r_\mathrm{in,2}$) with a gap between both components. The very narrow
disk width of 0.04 AU makes the inner disk region appear rather
ring-like in the NIR (see Fig.~\ref{figtgms}, right). We cannot
constrain the temperature gradient $q_1$ because the inner narrow
ring-shaped component is basically emitting only at one uniform
temperature $T_\mathrm{in,1}$. The inclination (angle between the
system axis and the viewing direction) is $48.6^{+2.9}_{-3.6}${\degr}
and the position angle of the disk is $169.3^{+4.2}_{-6.7}$\degr,
which is approximately consistent with our geometric model in
Sect.~\ref{kapgeo}. We emphasize that the structure of the outer disk
is a result of the longer wavelength data -- that is, the MIDI data
and the the MIR/far-infrared (FIR) SED. The inclination and position
angle of the system are mainly determined by the NIR interferometric
data but consistent with the MIDI data. In the case of
temperature-gradient models with more than one component, please note
that simultaneous modeling of the visibilities and the SED is able to
constrain the inner temperatures ($T_\mathrm{in,1}$,
$T_\mathrm{in,2}$) of the single components rather than the exact
shape of the single temperature gradients ($q_1$, $q_2$).

\begin{table*}[ht]
 \centering  
  \caption{Scanned range of the parameters of the computed
    temperature-gradient models (A, B$_1$, B$_2$ and B$_3$). }
  \label{tabpar}                     
  \begin{tabular}{lll|lllllll}  
    \hline \hline
    Parameter                    & Range (A)           & N(A)       & Range (B$_1$)       & N(B$_1$)     & Range (B$_2$)       & N(B$_2$)          & Range (B$_3$)       & N(B$_3$) \\ 
    \hline
    $r_\mathrm{in,1}$ [AU]        & 0.01 -- 1.0         & 30         & 0.05 -- 1.0         & 8            & 0.1 -- 0.3          & 8                 & 0.15 -- 0.25        & 10 \\
    $\Delta r_\mathrm{in,1}$ [AU] & 0.01 -- 10          & 30         & 0.01 -- 2.0         & 8            & 0.01 -- 0.15        & 8                 & 0.02 -- 0.07        & 8 \\
    $r_\mathrm{in,2}$ [AU]        & ...                 & ...        & 0.05 -- 3.0         & 8            & 0.4 -- 2.0          & 8                 & 1.0 -- 1.7          & 8 \\
    $\Delta r_\mathrm{in,2}$ [AU] & ...                 & ...        & 0.01 -- 10          & 8            & 0.5 -- 10           & 8                 & 1.0 -- 6.0          & 4 \\
    $T_\mathrm{in,1}$ [K]         & 200 -- 2000         & 30         & 800 -- 2000         & 8            & 1000 -- 1500        & 8                 & 1200 -- 1400        & 8 \\
    $T_\mathrm{in,2}$ [K]         & ...                 & ...        & 150 -- 1000         & 8            & 100 -- 400          & 8                 & 300 -- 380          & 8 \\
    $q_1$                        & 0.2, 0.5, 0.75, 1.0 & 4          & 0.2, 0.5, 0.75, 1.0 & 4            & 0.2, 0.5, 0.75, 1.0 & 4                 & 0.2, 0.5, 0.75, 1.0 & 4 \\
    $q_2$                        & ...                 & ...        & 0.2, 0.5, 0.75, 1.0 & 4            & 0.2, 0.5, 0.75, 1.0 & 4                 & 0.2, 0.5, 0.75, 1.0 & 4 \\
    $i$ [\degr]                  & 0 -- 90             & 30         & 0 -- 80             & 6            & 10 -- 60            & 6                 & 40 -- 60            & 8 \\
    $\vartheta$ [\degr]          & 0 -- 170            & 30         & 0 -- 160            & 6            & 130 -- 180          & 6                 & 150 -- 180          & 8 \\ 
    \hline
  \end{tabular}
  \tablefoot{The free model parameters for the inner ($j=1$) and, if
    included, outer ($j=2$) ring-shaped disk are as follows: the inner ring
    radius $r_\mathrm{in,j}$, the ring width $\Delta r_\mathrm{in,j}$,
    the temperature at the inner radius $T_\mathrm{in,j}$, the
    power-law index $q_j$, the inclination $i$, and the position angle
    of the semi-major axis $\vartheta$.  The number of steps computed
    per parameter is given by N. The step spacing is logarithmic for
    $r_\mathrm{in,1}$, $\Delta r_\mathrm{in,1}$, and
    $r_\mathrm{in,2}$, and is linear for the rest of the parameters
    except $q_1$ and $q_2$, for which the individual values used are
    given directly.}
\end{table*}

\begin{table}
  \centering  
  \caption{Parameters of best-fit temperature-gradient models A \&
    B$_3$.}
  \label{tabres}          
 \begin{tabular}{lll}  
    \hline \hline
    Parameter              & Best-fit            & Best-fit             \\ 
                           & parameters (A)      & parameters (B$_3$)   \\ \hline
    $r_\mathrm{in,1}$ [AU]  & $0.06 \pm 0.01$     & $0.19 \pm 0.01$      \\
    $r_\mathrm{out,1}$ [AU] & $4.6^{+1.1}_{-1.6}$  & $0.23 \pm 0.02$       \\
    $r_\mathrm{in,2}$ [AU]  & ...                 & $1.35^{+0.19}_{-0.20}$ \\
    $r_\mathrm{out,2}$ [AU] & ...                 & $>4.3$               \\ 
    $T_\mathrm{in,1}$ [K]   & $1937^{+25}_{-61}$   & $1257^{+133}_{-53}$    \\
    $T_\mathrm{in,2}$ [K]   & ...                 & $334^{+35}_{-17}$      \\
    $q_1$                  & $0.5$               & $0.5$                 \\
    $q_2$                  & ...                 & $1.0$                 \\
    $i$ [\degr]            & $86.9^{+0.2}_{-1.7}$  & $48.6^{+2.9}_{-3.6}$   \\
    $\vartheta$ [\degr]    & $139.0^{+0.2}_{-1.7}$ & $169.3^{+4.2}_{-6.7}$  \\ \hline
    $\chi_\mathrm{red}^2$   & 6.2                 & 2.7                   \\ \hline
  \end{tabular}
 \tablefoot{The listed parameters for the inner ($j=1$) and, if
   included, outer ($j=2$) ring-shaped disk are as follows: the inner
   ring radius $r_\mathrm{in,j}$, the outer ring radius
   $r_\mathrm{out,j}=r_\mathrm{in,j}+\Delta r_\mathrm{in,j}$, the
   temperature at the inner radius $T_\mathrm{in,j}$, the power-law
   index $q_j$, the inclination $i$, and the position angle of the
   semi-major axis $\vartheta$.  Parameters without error bars cannot
   be constrained (see discussion in Sect.~\ref{kapdis}).}
\end{table}

\section{Discussion} \label{kapdis}
\begin{figure}[tb]
  \includegraphics[angle=-90,width=0.5\textwidth]{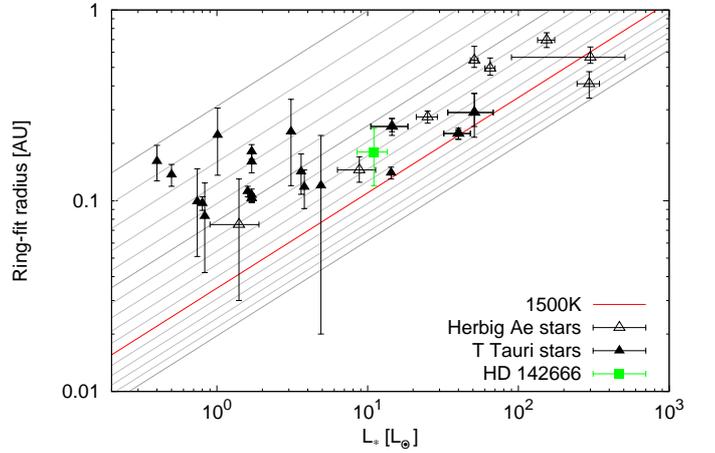}
  \caption{Size-luminosity diagram. The {\textit K}-band ring-fit
    radius of V1026~Sco (semi-major axis derived with an inclined ring
    fit in Sect.~\ref{kapgeo}) is plotted as a green filled
    square. For comparison, we also plot a sample of Herbig Ae stars
    (unfilled black triangles, \citealt{2005monmil}) and a sample of
    TTS (filled black triangles, \citealt{2008pinmen}). The
    theoretical relation between the ring radius in the NIR and the
    luminosity \citep{2002monmil} is shown for different
    temperatures. The 1500~K line is plotted in red; the gray lines
    indicate curves for temperatures between 500~K (highest line) and
    2000~K (lowest line) in 100~K steps.}
  \label{figslr}
\end{figure}

To compare the disk size of V1026~Sco with other young stellar
objects, we plot the {\textit K}-band radius ($0.18 \pm 0.06$~AU, see
Table~\ref{tabrad}) obtained with a geometric inclined-ring fit into
the size-luminosiy diagram (Fig.~\ref{figslr}). The derived radius is
$\sim$$1.5$ times larger than expected for a dust sublimation
temperature of 1500~K and is located approximately on the 1200~K
line. Within the error bars, it still agrees with a dust sublimation
temperature of 1500~K expected for dust consisting mostly of silicates
\citep{2001natpru,2001duldom,2002monmil}.

The derived best-fit temperature-gradient model B$_3$ consists of a
two-component disk model; all parameters are listed in
Table~\ref{tabres}. The inner component is a ring-shaped disk with an
inner radius of $0.19 \pm 0.01$~AU and an inner temperature of
$1257^{+133}_{-53}$~K. The outer ring-shaped disk extends from
$1.35^{+0.19}_{-0.20}$~AU to $>$~4.3~AU with $334^{+35}_{-17}$~K at
the inner edge. Between the hot inner and cool outer components, our
best-fit model shows an approximately 1.1~AU-wide gap. The inclination
of $48.6^{+2.9}_{-3.6}${\degr} is similar to the inclination derived
with our geometric model in Sect.~\ref{kapgeo} ($\sim$49\degr and
$\sim$50\degr) and to the value of 55{\degr} found by
\citet{2003domdul}. Neither of the power-law indices could be
constrained. In the inner component, the ring width is too small to
allow us to constrain $q_1$. In the outer component, there seems to be
a degeneracy between $q_2$ and $r_\mathrm{out,2}$. The parameters of
the inner disk (Table~\ref{tabres}) are similar to the ones obtained
with the geometric fit of the near-infrared visibilities
(Table~\ref{tabrad}) and the size-luminosity diagram
(Fig.~\ref{figslr}).
A possible explanation for the lack of FIR emission in the SED of some
HAeBes and the apparent disk gap is self-shadowing by the puffed-up
inner rim \citep{2004duldom}.

The stellar rotation of V1026~Sco is $v \sin i=65.3 \pm
3.1$~km~s$^{-1}$ \citep{2013alewad}, leading to a maximum rotational
velocity of $\sim$$86$~km~s$^{-1}$ if the inclination of the star is
comparable to the disk inclination. This is similar \citep{1995boecat}
or slightly lower \citep{2013alewadb} than the average velocity of
low-mass Herbig Ae/Be stars but higher than measurements of T~Tauri
stars \citep{2010weilau}.

Magnetic braking can reduce the rotational velocity as observed in
T~Tauri stars \citep{1991koe,2010weilau,2013johjar}. In Herbig stars,
strong magnetic braking is less likely than in T~Tauri stars because
the observed magnetic fields are weaker. In V1026~Sco, a magnetic
field has not been detected \citep{2013alewad}, which could mean that
it is weak, as expected for Herbig Ae stars
\citep{2010weilau}. Even if a magnetic field exists, but could not be
detected, the high value of the stellar rotation velocity of V1026~Sco
suggests that rotational braking via disk locking
\citep{1991koe,2000ste} is much weaker than in T~Tauri stars.

Our findings can be described with the standard disk theories for
Herbig Ae stars, which postulate passive circumstellar disks with
inner holes and puffed-up inner rims \citep{2001natpru,2001duldom}. In
addition, the derived inclination might be large enough to explain the
UXOr variability of V1026~Sco in the context of current proposed
theories, as we discuss in the following. Theories about partial
obscuration of the stellar light by hydrodynamic fluctuations of the
inner rim need high inclination angles for explaining the UXOr
variability \citep{2003dulvan}. With the measured intermediate
inclination of V1026~Sco, the rim fluctuations would have to be twice
as high as the theoretical fluctuation height. Therefore, this model
is less suitable in our case and also for several other UXOrs
(e.g. \citealt{2007pondul}). The unsteady accretion model by
\citet{1999hershe} is inclination independent and cannot be disproven
with our measurements. For intermediate to high disk inclinations,
orbiting dust clouds might intercept the line of sight toward the star
\citep{1994grithe,1997natgri}. For this case, the derived inclination
of V1026~Sco is still within the range predicted for UXOrs
(45\degr--68\degr) by \citet{2000natwhi}. Dust clouds in
centrifugally-driven disk winds \citep{2007vinjur,2012bankoe} can also
explain the UX~Ori type variability of V1026~Sco, as they also are
consistent with intermediate to high disk inclinations.

\section{Conclusion}
We observed the UX~Ori star V1026~Sco with VLTI/AMBER in the
\textit{H} and \textit{K} bands. With a geometric ring-shaped model
consisting of the star and an inclined ring, we found a radius of
$r_\mathrm{ring,in}=0.18 \pm 0.06$~AU in the \textit{K} band. In the
context of the size-luminosity diagram, this radius is found to be
consistent with the theory of a passive circumstellar disk with an
inner hole and a rim at the dust sublimation radius.  We further
derived an inclination of $50 \pm 11\degr$ and $49 \pm 5\degr$ and a
PA of the semi-major axis of the inclined disk of $179 \pm 17\degr$
and $163 \pm 9\degr$ in the \textit{H} and \textit{K} bands,
respectively.

We found a two-component-disk temperature-gradient model that is able
to reproduce all visibilities and the SED. The inner radius of the
inner disk is $0.19 \pm 0.01$~AU and similar to the one found with a
geometric ring fit. The two disk components are separated by a gap,
which may be explained by a shadow cast by a puffed-up inner rim and
agrees with the type II classification of the object. The derived
inclination of $48.6^{+2.9}_{-3.6}${\degr} and the PA of
$169.3^{+4.2}_{-6.7}${\degr} are consistent with the values found by
geometric modeling. Our inclination of $\sim 49\degr$ is probably not
consistent with a model where rim fluctuations cause the UXOr
variability, because the expected rim height is not high enough, as
discussed above. The unsteady accretion theory cannot be excluded with
our measurements, because the model is
inclination-independent. Finally, the measured intermediate disk
inclination is within the range predicted from UXOr models with
orbiting dust clouds in the disk or in centrifugally-driven disk
winds.


\begin{acknowledgements}
  We would like to thank K. R. W. Tristram for helpful discussions and
  suggestions and also our colleagues at Paranal for their excellent
  collaboration. We thank the anonymous referees for the helpful
  comments.
\end{acknowledgements}

\bibliographystyle{aa} 
\bibliography{/aux/pc20218a/jvural/BIBO}

\begin{thebibliography}{41}
\expandafter\ifx\csname natexlab\endcsname\relax\def\natexlab#1{#1}\fi

\bibitem[{{Alecian} {et~al.}(2013{\natexlab{a}}){Alecian}, {Wade}, {Catala},
  {Grunhut}, {Landstreet}, {Bagnulo}, {B{\"o}hm}, {Folsom}, {Marsden}, \&
  {Waite}}]{2013alewad}
{Alecian}, E., {Wade}, G.~A., {Catala}, C., {et~al.} 2013{\natexlab{a}},
  \mnras, 429, 1001

\bibitem[{{Alecian} {et~al.}(2013{\natexlab{b}}){Alecian}, {Wade}, {Catala},
  {Grunhut}, {Landstreet}, {B{\"o}hm}, {Folsom}, \& {Marsden}}]{2013alewadb}
{Alecian}, E., {Wade}, G.~A., {Catala}, C., {et~al.} 2013{\natexlab{b}},
  \mnras, 429, 1027

\bibitem[{{Bans} \& {K{\"o}nigl}(2012)}]{2012bankoe}
{Bans}, A. \& {K{\"o}nigl}, A. 2012, \apj, 758, 100

\bibitem[{{Boehm} \& {Catala}(1995)}]{1995boecat}
{Boehm}, T. \& {Catala}, C. 1995, \aap, 301, 155

\bibitem[{{Bord{\'e}} {et~al.}(2002){Bord{\'e}}, {Coud{\'e} du Foresto},
  {Chagnon}, \& {Perrin}}]{2002borcou}
{Bord{\'e}}, P., {Coud{\'e} du Foresto}, V., {Chagnon}, G., \& {Perrin}, G.
  2002, \aap, 393, 183

\bibitem[{{Chelli} {et~al.}(2009){Chelli}, {Utrera}, \& {Duvert}}]{2009cheutr}
{Chelli}, A., {Utrera}, O.~H., \& {Duvert}, G. 2009, \aap, 502, 705

\bibitem[{{Dominik} {et~al.}(2003){Dominik}, {Dullemond}, {Waters}, \&
  {Walch}}]{2003domdul}
{Dominik}, C., {Dullemond}, C.~P., {Waters}, L.~B.~F.~M., \& {Walch}, S. 2003,
  \aap, 398, 607

\bibitem[{{Dullemond} \& {Dominik}(2004)}]{2004duldom}
{Dullemond}, C.~P. \& {Dominik}, C. 2004, \aap, 417, 159

\bibitem[{{Dullemond} {et~al.}(2001){Dullemond}, {Dominik}, \&
  {Natta}}]{2001duldom}
{Dullemond}, C.~P., {Dominik}, C., \& {Natta}, A. 2001, \apj, 560, 957

\bibitem[{{Dullemond} {et~al.}(2003){Dullemond}, {van den Ancker}, {Acke}, \&
  {van Boekel}}]{2003dulvan}
{Dullemond}, C.~P., {van den Ancker}, M.~E., {Acke}, B., \& {van Boekel}, R.
  2003, \apjl, 594, L47

\bibitem[{{Folsom} {et~al.}(2012){Folsom}, {Bagnulo}, {Wade}, {Alecian},
  {Landstreet}, {Marsden}, \& {Waite}}]{2012folbag}
{Folsom}, C.~P., {Bagnulo}, S., {Wade}, G.~A., {et~al.} 2012, \mnras, 422, 2072

\bibitem[{{Grinin} {et~al.}(2001){Grinin}, {Kozlova}, {Natta}, {Ilyin},
  {Tuominen}, {Rostopchina}, \& {Shakhovskoy}}]{2001grikoz}
{Grinin}, V.~P., {Kozlova}, O.~V., {Natta}, A., {et~al.} 2001, \aap, 379, 482

\bibitem[{{Grinin} {et~al.}(1994){Grinin}, {The}, {de Winter}, {Giampapa},
  {Rostopchina}, {Tambovtseva}, \& {van den Ancker}}]{1994grithe}
{Grinin}, V.~P., {The}, P.~S., {de Winter}, D., {et~al.} 1994, \aap, 292, 165

\bibitem[{{Herbst} \& {Shevchenko}(1999)}]{1999hershe}
{Herbst}, W. \& {Shevchenko}, V.~S. 1999, \aj, 118, 1043

\bibitem[{{Johnstone} {et~al.}(2013){Johnstone}, {Jardine}, {Gregory},
  {Donati}, \& {Hussain}}]{2013johjar}
{Johnstone}, C.~P., {Jardine}, M., {Gregory}, S.~G., {Donati}, J.-F., \&
  {Hussain}, G. 2013, \mnras

\bibitem[{{Juh{\'a}sz} {et~al.}(2010){Juh{\'a}sz}, {Bouwman}, {Henning},
  {Acke}, {van den Ancker}, {Meeus}, {Dominik}, {Min}, {Tielens}, \&
  {Waters}}]{2010juhbou}
{Juh{\'a}sz}, A., {Bouwman}, J., {Henning}, T., {et~al.} 2010, \apj, 721, 431

\bibitem[{{Kishimoto} {et~al.}(2011){Kishimoto}, {H{\"o}nig}, {Antonucci},
  {Millour}, {Tristram}, \& {Weigelt}}]{2011kishoe}
{Kishimoto}, M., {H{\"o}nig}, S.~F., {Antonucci}, R., {et~al.} 2011, \aap, 536,
  A78

\bibitem[{{Koenigl}(1991)}]{1991koe}
{Koenigl}, A. 1991, \apjl, 370, L39

\bibitem[{{Kreplin} {et~al.}(2012){Kreplin}, {Kraus}, {Hofmann}, {Schertl},
  {Weigelt}, \& {Driebe}}]{2012krekra}
{Kreplin}, A., {Kraus}, S., {Hofmann}, K.-H., {et~al.} 2012, \aap, 537, A103

\bibitem[{{Lafrasse} {et~al.}(2010){Lafrasse}, {Mella}, {Bonneau}, {Duvert},
  {Delfosse}, \& {Chelli}}]{2010lafmel}
{Lafrasse}, S., {Mella}, G., {Bonneau}, D., {et~al.} 2010, VizieR Online Data
  Catalog, 2300, 0

\bibitem[{{Lecavelier des Etangs} {et~al.}(2005){Lecavelier des Etangs},
  {Nitschelm}, {Olsen}, {Vidal-Madjar}, \& {Ferlet}}]{2005lecnit}
{Lecavelier des Etangs}, A., {Nitschelm}, C., {Olsen}, E.~H., {Vidal-Madjar},
  A., \& {Ferlet}, R. 2005, \aap, 439, 571

\bibitem[{{Leinert} {et~al.}(2003){Leinert}, {Graser}, {Przygodda}, {Waters},
  {Perrin}, {Jaffe}, {Lopez}, {Bakker}, {B{\"o}hm}, {Chesneau}, {Cotton},
  {Damstra}, {de Jong}, {Glazenborg-Kluttig}, {Grimm}, {Hanenburg}, {Laun},
  {Lenzen}, {Ligori}, {Mathar}, {Meisner}, {Morel}, {Morr}, {Neumann}, {Pel},
  {Schuller}, {Rohloff}, {Stecklum}, {Storz}, {von der L{\"u}he}, \&
  {Wagner}}]{2003leigra}
{Leinert}, C., {Graser}, U., {Przygodda}, F., {et~al.} 2003, \apss, 286, 73

\bibitem[{{Meeus} {et~al.}(1998){Meeus}, {Waelkens}, \& {Malfait}}]{1998meewae}
{Meeus}, G., {Waelkens}, C., \& {Malfait}, K. 1998, \aap, 329, 131

\bibitem[{{Mendigut{\'{\i}}a} {et~al.}(2011){Mendigut{\'{\i}}a}, {Calvet},
  {Montesinos}, {Mora}, {Muzerolle}, {Eiroa}, {Oudmaijer}, \&
  {Mer{\'{\i}}n}}]{2011mencal}
{Mendigut{\'{\i}}a}, I., {Calvet}, N., {Montesinos}, B., {et~al.} 2011, \aap,
  535, A99

\bibitem[{{Monnier} \& {Millan-Gabet}(2002)}]{2002monmil}
{Monnier}, J.~D. \& {Millan-Gabet}, R. 2002, \apj, 579, 694

\bibitem[{{Monnier} {et~al.}(2005){Monnier}, {Millan-Gabet}, {Billmeier},
  {Akeson}, {Wallace}, {Berger}, {Calvet}, {D'Alessio}, {Danchi}, {Hartmann},
  {Hillenbrand}, {Kuchner}, {Rajagopal}, {Traub}, {Tuthill}, {Boden}, {Booth},
  {Colavita}, {Gathright}, {Hrynevych}, {Le Mignant}, {Ligon}, {Neyman},
  {Swain}, {Thompson}, {Vasisht}, {Wizinowich}, {Beichman}, {Beletic},
  {Creech-Eakman}, {Koresko}, {Sargent}, {Shao}, \& {van Belle}}]{2005monmil}
{Monnier}, J.~D., {Millan-Gabet}, R., {Billmeier}, R., {et~al.} 2005, \apj,
  624, 832

\bibitem[{{Natta} {et~al.}(1997){Natta}, {Grinin}, {Mannings}, \&
  {Ungerechts}}]{1997natgri}
{Natta}, A., {Grinin}, V.~P., {Mannings}, V., \& {Ungerechts}, H. 1997, \apj,
  491, 885

\bibitem[{{Natta} {et~al.}(2001){Natta}, {Prusti}, {Neri}, {Wooden}, {Grinin},
  \& {Mannings}}]{2001natpru}
{Natta}, A., {Prusti}, T., {Neri}, R., {et~al.} 2001, \aap, 371, 186

\bibitem[{{Natta} \& {Whitney}(2000)}]{2000natwhi}
{Natta}, A. \& {Whitney}, B.~A. 2000, \aap, 364, 633

\bibitem[{{Petrov} {et~al.}(2007){Petrov}, {Malbet}, {Weigelt}, {Antonelli},
  {Beckmann}, {Bresson}, {Chelli}, {Dugu{\'e}}, {Duvert}, {Gennari},
  {Gl{\"u}ck}, {Kern}, {Lagarde}, {Le Coarer}, {Lisi}, {Millour}, {Perraut},
  {Puget}, {Rantakyr{\"o}}, {Robbe-Dubois}, {Roussel}, {Salinari}, {Tatulli},
  {Zins}, {Accardo}, {Acke}, {Agabi}, {Altariba}, {Arezki}, {Aristidi},
  {Baffa}, {Behrend}, {Bl{\"o}cker}, {Bonhomme}, {Busoni}, {Cassaing},
  {Clausse}, {Colin}, {Connot}, {Delboulb{\'e}}, {Domiciano de Souza},
  {Driebe}, {Feautrier}, {Ferruzzi}, {Forveille}, {Fossat}, {Foy},
  {Fraix-Burnet}, {Gallardo}, {Giani}, {Gil}, {Glentzlin}, {Heiden},
  {Heininger}, {Hernandez Utrera}, {Hofmann}, {Kamm}, {Kiekebusch}, {Kraus},
  {Le Contel}, {Le Contel}, {Lesourd}, {Lopez}, {Lopez}, {Magnard}, {Marconi},
  {Mars}, {Martinot-Lagarde}, {Mathias}, {M{\`e}ge}, {Monin}, {Mouillet},
  {Mourard}, {Nussbaum}, {Ohnaka}, {Pacheco}, {Perrier}, {Rabbia}, {Rebattu},
  {Reynaud}, {Richichi}, {Robini}, {Sacchettini}, {Schertl}, {Sch{\"o}ller},
  {Solscheid}, {Spang}, {Stee}, {Stefanini}, {Tallon}, {Tallon-Bosc}, {Tasso},
  {Testi}, {Vakili}, {von der L{\"u}he}, {Valtier}, {Vannier}, \&
  {Ventura}}]{2007petmal}
{Petrov}, R.~G., {Malbet}, F., {Weigelt}, G., {et~al.} 2007, \aap, 464, 1

\bibitem[{{Pinte} {et~al.}(2008){Pinte}, {M{\'e}nard}, {Berger}, {Benisty}, \&
  {Malbet}}]{2008pinmen}
{Pinte}, C., {M{\'e}nard}, F., {Berger}, J.~P., {Benisty}, M., \& {Malbet}, F.
  2008, \apjl, 673, L63

\bibitem[{{Pontoppidan} {et~al.}(2007){Pontoppidan}, {Dullemond}, {Blake},
  {Boogert}, {van Dishoeck}, {Evans}, {Kessler-Silacci}, \&
  {Lahuis}}]{2007pondul}
{Pontoppidan}, K.~M., {Dullemond}, C.~P., {Blake}, G.~A., {et~al.} 2007, \apj,
  656, 980

\bibitem[{{Richichi} {et~al.}(2005){Richichi}, {Percheron}, \&
  {Khristoforova}}]{2005ricper}
{Richichi}, A., {Percheron}, I., \& {Khristoforova}, M. 2005, \aap, 431, 773

\bibitem[{{Schegerer} {et~al.}(2013){Schegerer}, {Ratzka}, {Schuller}, {Wolf},
  {Mosoni}, \& {Leinert}}]{2013schrat}
{Schegerer}, A.~A., {Ratzka}, T., {Schuller}, P.~A., {et~al.} 2013, \aap, 555,
  A103

\bibitem[{{St{\c e}pie{\'n}}(2000)}]{2000ste}
{St{\c e}pie{\'n}}, K. 2000, \aap, 353, 227

\bibitem[{{Tatulli} {et~al.}(2007){Tatulli}, {Millour}, {Chelli}, {Duvert},
  {Acke}, {Hernandez Utrera}, {Hofmann}, {Kraus}, {Malbet}, {M{\`e}ge},
  {Petrov}, {Vannier}, {Zins}, {Antonelli}, {Beckmann}, {Bresson}, {Dugu{\'e}},
  {Gennari}, {Gl{\"u}ck}, {Kern}, {Lagarde}, {Le Coarer}, {Lisi}, {Perraut},
  {Puget}, {Rantakyr{\"o}}, {Robbe-Dubois}, {Roussel}, {Weigelt}, {Accardo},
  {Agabi}, {Altariba}, {Arezki}, {Aristidi}, {Baffa}, {Behrend}, {Bl{\"o}cker},
  {Bonhomme}, {Busoni}, {Cassaing}, {Clausse}, {Colin}, {Connot},
  {Delboulb{\'e}}, {Domiciano de Souza}, {Driebe}, {Feautrier}, {Ferruzzi},
  {Forveille}, {Fossat}, {Foy}, {Fraix-Burnet}, {Gallardo}, {Giani}, {Gil},
  {Glentzlin}, {Heiden}, {Heininger}, {Kamm}, {Kiekebusch}, {Le Contel}, {Le
  Contel}, {Lesourd}, {Lopez}, {Lopez}, {Magnard}, {Marconi}, {Mars},
  {Martinot-Lagarde}, {Mathias}, {Monin}, {Mouillet}, {Mourard}, {Nussbaum},
  {Ohnaka}, {Pacheco}, {Perrier}, {Rabbia}, {Rebattu}, {Reynaud}, {Richichi},
  {Robini}, {Sacchettini}, {Schertl}, {Sch{\"o}ller}, {Solscheid}, {Spang},
  {Stee}, {Stefanini}, {Tallon}, {Tallon-Bosc}, {Tasso}, {Testi}, {Vakili},
  {von der L{\"u}he}, {Valtier}, \& {Ventura}}]{2007tatmil}
{Tatulli}, E., {Millour}, F., {Chelli}, A., {et~al.} 2007, \aap, 464, 29

\bibitem[{{van Boekel} {et~al.}(2005){van Boekel}, {Min}, {Waters}, {de Koter},
  {Dominik}, {van den Ancker}, \& {Bouwman}}]{2005vanmin}
{van Boekel}, R., {Min}, M., {Waters}, L.~B.~F.~M., {et~al.} 2005, \aap, 437,
  189

\bibitem[{{Vinkovi{\'c}} \& {Jurki{\'c}}(2007)}]{2007vinjur}
{Vinkovi{\'c}}, D. \& {Jurki{\'c}}, T. 2007, \apj, 658, 462

\bibitem[{{Vural} {et~al.}(2012){Vural}, {Kreplin}, {Kraus}, {Weigelt},
  {Driebe}, {Benisty}, {Dugu{\'e}}, {Massi}, {Monin}, \&
  {Vannier}}]{2012vurkre}
{Vural}, J., {Kreplin}, A., {Kraus}, S., {et~al.} 2012, \aap, 543, A162

\bibitem[{{Weise} {et~al.}(2010){Weise}, {Launhardt}, {Setiawan}, \&
  {Henning}}]{2010weilau}
{Weise}, P., {Launhardt}, R., {Setiawan}, J., \& {Henning}, T. 2010, \aap, 517,
  A88

\bibitem[{{Zwintz} {et~al.}(2009){Zwintz}, {Kallinger}, {Guenther},
  {Gruberbauer}, {Huber}, {Rowe}, {Kuschnig}, {Weiss}, {Matthews}, {Moffat},
  {Rucinski}, {Sasselov}, {Walker}, \& {Casey}}]{2009zwikal}
{Zwintz}, K., {Kallinger}, T., {Guenther}, D.~B., {et~al.} 2009, \aap, 494,
  1031

\end{thebibliography}

\end{document}